\documentclass[english]{llncs}

\usepackage{amssymb}
\usepackage{amsmath}

\usepackage{multirow}

\usepackage{graphicx}
\usepackage{subfigure}

\newcommand{\locations}{\ensuremath{V}}

\newcommand{\loadd}{\ensuremath{x}}
\newcommand{\driver}{\ensuremath{driv}}
\newcommand{\ndriver}{\ensuremath{k}}
\newcommand{\orig}{\ensuremath{orig}}
\newcommand{\origin}{\ensuremath{orig}}
\newcommand{\dest}{\ensuremath{dest}}
\newcommand{\tdep}{\ensuremath{dep}}
\newcommand{\tarr}{\ensuremath{arr}}
\newcommand{\mloadd}{\ensuremath{\ell oad}}

\newcommand{\ntourd}{\ensuremath{n}}
\newcommand{\taskset}{\ensuremath{\mathcal{T}}}

\newcommand{\abs}[1]{\ensuremath{\lvert #1 \rvert}}

\newcommand{\NN}{\ensuremath{\mathbb{N}}}
\newcommand{\ZZ}{\ensuremath{\mathbb{Z}}}

\newcommand{\RR}{\ensuremath{\mathbb{R}}}

\newcommand{\capd}{\ensuremath{L}}
\newcommand{\tourd}{\ensuremath{\Gamma}}
\newcommand{\ptourd}{\ensuremath{\check{\Gamma}}}

\newcommand{\move}{\ensuremath{m}}
\newcommand{\pmove}{{\ensuremath{\check{m}}}}
\newcommand{\dist}{\ensuremath{d}}

\newcommand{\fc}{\ensuremath{f}}
\newcommand{\fd}{\ensuremath{F}}

\newcommand{\cd}{\ensuremath{C}}

\newcommand{\bc}{\ensuremath{b}}

\newcommand{\capacity}{\ensuremath{\text{cap}}}

\newcommand{\sched}{\mathcal{S}}

\newcommand{\z}{\boldsymbol{z}}

\newcommand{\NPhard}{\ensuremath{\mathcal{NP}}-hard}
\newcommand{\LiftFlow}{\textsc{LiftFlow}}
\newcommand{\REOPT}{\textsc{ReOpt}}

\allowdisplaybreaks[4]

\title{Two Flow-Based Approaches for 
the Static Relocation Problem in Carsharing Systems}
\titlerunning{Two Flow-Based Approaches for 
the Static Relocation Problem in Carsharing Systems}

\author{Sahar Bsaybes${}^\star$
        \and
        Alain Quilliot
        \and
        Annegret K.\ Wagler
        \and
        Jan-Thierry Wegener\thanks{This work was founded by the French National Research Agency, the European Commission (Feder funds) and the R\'egion Auvergne within the LabEx IMobS3.}}
\institute{Universit\'e Blaise Pascal (Clermont-Ferrand II)\\
          Laboratoire d'Informatique, de Mod\'elisation et d'Optimisation des Syst\`emes\\
          BP 10125, 63173 Aubi\`ere Cedex, France\\
        \texttt{\{bsaybes,quilliot,wagler,wegener\}@isima.fr}}

\begin{document}

\maketitle

\begin{abstract}
In a carsharing system, a fleet of cars is distributed at stations in an urban area, customers can take and return cars at any time and station.
For operating such a system in a satisfactory way, the stations have to keep a good ratio between the numbers of free places and cars in each station.
This leads to the problem of relocating cars between stations, which can be modeled within the framework of a metric task system. 
In this paper, we focus on the Static Relocation Problem, where the system has to be set into a certain state, outgoing from the current state.  
We present two approaches to solve this problem, a fast heuristic approach and an exact integer programming based method using flows in time-expanded networks, and provide some computational results.
\end{abstract}

\section{Introduction}

Carsharing is a modern way of car rental, attractive to customers who make only occasional use of a car on demand.
In a carsharing system, a fleet of cars is distributed at specified stations in an urban area, customers can take a car at any time and station
and return it at any time and station, provided that there is a car available at the start station and a free place at the final station.
To ensure the latter, the stations have to keep a good ratio between the number of places and the number of cars in each station.
This leads to the problem of balancing the load of the stations, called \emph{Relocation Problem}:
an operator has to monitor the load situations of the stations and to decide when and how to move cars from ``overfull'' stations to ``underfull'' ones.

Balancing problems of this type occur for any car- or bikesharing system, but the scale of the instances 
and the possibility to move one or more vehicles in balancing steps differ.
We consider an innovative carsharing system, where the cars are partly autonomous, which allows to build wireless convoys of cars leaded by a special vehicle,
such that the whole convoy is moved by only one driver (cf.~\cite{EDGC:2012:PCS}).
This setting is similar to bikesharing, where trucks can simultaneously move several bikes during the relocation process \cite{Benchimol+etal:RAIRO,do-cmc2013,cirrelt-CMR-2012}.
The main goal is to guarantee a balanced system during working hours (dynamic situation as in \cite{cirrelt-CMR-2012,KQWW:2014:LNCS})
or to set up an appropriate initial state for the morning (static situation as in \cite{do-cmc2013}).
Both, the dynamic and the static versions are known to be \NPhard~\cite{Ball+etal:handbook:95a,do-cmc2013}, 
a combinatorial approximation algorithm \REOPT\ \cite{LAGOS2013} and different heuristic approaches have been developed, see e.g. \cite{KQWW:2014:LNCS,EvoCOP:HPHR-2013,SHH-2013}.

In this paper, we address the static situation where the system has to be set into a certain state, outgoing from the current state, within a given time horizon (Static Relocation Problem).
In Section~\ref{Sec_Modeling}, we model the problem within the framework of a metric task system, where the tasks consist in moving cars out of “overfull” stations into “underfull” ones.
Then, we present an exact and a heuristic approach to solve this problem.
In order to obtain an exact solution, we interpret the Static Relocation Problem by means of flows in a time-expanded network (as e.g., proposed by~\cite{do-cmc2013} for bikesharing systems),
see Section~\ref{sec: ten}.
Hereby, the two flows are not independent or share arc capacities as in the case of multicommodity flows, but are coupled in the sense that the flow of cars is dependent from the flow of drivers (since cars can only be moved in convoys).
As obtaining an exact solution requires long computation times 
and the approximation algorithm from \cite{LAGOS2013} does not always find a feasible solution, 
we propose a heuristic in Section~\ref{sec: aggregated network}, which firstly computes coupled flows in an aggregated network, and afterwards generates tours from these flows by time-expansion. 
We show that this approach yields optimal solutions under certain conditions and finally provide some computational results for both approaches 
in comparison with a lower bound and the approximation algorithm from \cite{LAGOS2013}, 
and close with some remarks and future lines of research.

\section{Problem Description and Model}
\label{Sec_Modeling}

We model the Relocation Problem in the framework of a metric task system. 

By \cite{LAGOS2013}, the studied carsharing system can be understood as a discrete event-based system, where
the system components are the stations $v_1, \ldots, v_n$, each having an individual capacity cap$(v_i)$,
a system state $\z^t \in \ZZ^n$ specifies for each station $v_i$ the number of cars $z^t_i$ at a time point $t \leq T$ within a time horizon~$[0, T]$ 
and $\z^t$ changes when customers or convoy drivers take or return cars at stations.

An operator monitors the evolution of system states over time and decides when and how to move cars from overfull stations to underfull ones, in order to avoid infeasible system states $\z^t$ with $z^t_i >$ cap$(v_i)$ or $z^t_i<0$ for some station $v_i$. 
More precisely, a \emph{task} is defined by $\tau = (v_i, x)$ where $x \in \ZZ \setminus \{0\}$ cars are to pickup (if $x > 0$) or to deliver (if $x < 0$) at station $v_i$ within the time-horizon $[0, T]$. 
We call a task \emph{oversatisfied} if more than $\abs{x}$~cars are picked up (resp.~dropped) at $v_i$.

To fulfill these tasks, we create tours for the convoys in order to perform the desired relocation process.
For that, it is suitable to encode the urban area where the carsharing system is running as a \emph{metric space} $M=(V,d)$ induced by a weighted graph $G = (V,E)$ with weight function $w : E \rightarrow \RR_+$, where 
the nodes correspond to stations, 
edges to their physical links in the urban area, and the
distance $d$ between two points $v_i,v_j \in V$ is the length of a shortest path from $v_i$ to $v_j$. 
All drivers begin and end their work at the same location, the so-called depot.
The depot is represented in $V$ by a distinguished origin $v_0 \in V$.

This together yields a \emph{metric task system}, a pair $(M, {\cal T})$ where $M = (V, d)$ is the above metric space and ${\cal T}$ a set of tasks, as suitable framework to embed the tours for the convoys. 
A driver able to lead a convoy plays the role of a server, the number of available drivers is denoted by~$k$.
Each server has capacity~$L$, corresponding to the maximum possible number of cars per convoy; several servers are necessary to serve a task $\tau = (v, x)$ if $x > L$ holds.

More precisely, we define the following.
A \emph{move} from one station to another is
$\move = (j, v, t_v, v', t_{v'}, \loadd)$, where
$j \in \{ 1, \dotsc, \ndriver \}$ specifies the driver $\driver(\move)$ that has to move from the origin station $\origin(\move) = v \in \locations$ starting at time $\tdep(\move) = t_v$
to destination station $\dest(\move) = v' \in \locations$ arriving at time $\tarr(\move) = t_{v'}$, and a load of $\mloadd(\move) = \loadd$ cars.
Hereby, we require that 
\begin{itemize}
 \item\label{def: enum: move: 3} $0 \leq \mloadd(\move_i) \leq \capd$,
 \item\label{def: enum: move: 2} from $\orig(\move) \neq \dest(\move)$ follows $\tarr(\move) = \tdep(\move) + \dist(\orig(\move), \dest(\move))$.
\end{itemize}
We speak about a \emph{waiting move} if $\orig(\move) = \dest(\move)$ holds (which might be necessary if a proceeding move cannot be performed before a certain time point).

A \emph{tour} is a sequence $\tourd = (\move_1, \move_2, \dotsc, \move_\ntourd)$ of moves, starting and ending in the depot, with
\begin{itemize}
 \item\label{def: enum: tour: 1} $\driver(\move_1) = \dotsm = \driver(\move_\ntourd)$,
\item\label{def: enum: tour: 2} $\dest(\move_i) = \orig(\move_{i+1})$,
 \item\label{def: enum: tour: 3} $\tarr(\move_i) = \tdep(\move_{i+1})$.
\end{itemize}%

A move~$\move_i$ \emph{picks up} cars at~$\orig(\move_i)$, if $\mloadd(\move_{i-1}) - \mloadd(\move_i) < 0$,
and $\move_i$ \emph{drops} cars at~$\dest(\move_i)$, if $\mloadd(\move_i) - \mloadd(\move_{i-1}) < 0$.
The \emph{length of a tour} corresponds to the distance traveled by the driver.
Several tours are composed to a transportation schedule. 
A collection of tours $\{\tourd_1, \ldots, \tourd_\ndriver \}$ is a \emph{feasible transportation schedule} $\sched$ for $(M, \taskset)$ if 
\begin{enumerate}
 \item\label{def: enum: schedule: 1} every driver has exactly one tour,
 \item\label{def: enum: schedule: 2} each task $\tau \in \taskset$ is served (i.e., for every task $\tau = (v, x)$, the number of cars picked up (resp.~dropped) at station~$v$ sum up to~$x$,
 \item\label{def: enum: schedule: 3} all system states $\z^t$ are feasible during the whole time horizon $[0, T]$. 
\end{enumerate}
The \emph{total tour length} of a transportation schedule is the sum of the lengths of its tours.
Condition~\ref{def: enum: schedule: 3} requires that, besides the canonical precedences between a move $\move_i \in \tourd$ and its successor move $\move_{i+1} \in \tourd$, also dependencies between tours are respected
if \emph{preemption} is used, i.e., if a car is transported in one convoy from its origin to an intermediate station, and from there by another convoy to its destination.
This causes dependencies between tours, since some moves cannot be performed before others are done without leading to infeasible intermediate states (the reason why tours may contain waiting moves).
More precisely, there is a precedence between move $\move_i \in \tourd$ and move $\move'_{j} \in \tourd'$ avoiding a system state $z^t$ with $z^t_v < 0$ (resp.~$cap(v) < z^t_v$),
if one of the following conditions is true:
\begin{itemize}
\item the move $\move_i$ drops cars at an overfull station,
\item the move $\move_j$ picks up cars at underfull station.
\end{itemize}

We call a transportation schedule \emph{non-preemptive} if there are no precedences between moves of different tours, and \emph{preemptive} otherwise.

Our goal is to construct transportation schedules of minimal total tour length for the Relocation Problem in the static situation; hereby, all tours have to start and to end in the depot $v_0$, and preemption is allowed.

\begin{problem}[Static Relocation Problem $(M,\z^0,\z^T,k,L)$]
Given a metric space $M=(V, \dist)$ induced by a weighted graph $G=(V, E, w)$,
start state $\z^0 \in \NN^{|V|}$, destination state $\z^T \in \NN^{|V|}$ with $\abs{\z^0} = \abs{\z^T}$ and time horizon $T$, $k$ servers of capacity $L$,
find a transportation schedule of minimal total tour length for the metric task system $(M, {\cal T})$ where ${\cal T}$ consists of the tasks $\tau = (v_i, z_i^0-z_i^T)$ for all~$v_i$ with $z_i^0 \neq z_i^T$.
\end{problem}

\section{Min-Cost Flows in Time-Expanded Networks}
\label{sec: ten}

In this section, we give an exact approach for the Static Relocation Problem $(M,\z^0,\z^T,k,L)$ by defining a time-expanded network with two coupled flows.
For that, we build a directed graph~$G^T = (V_T, A_T)$, with $A_T = A_H \cup A_L$, as a time-expanded version of the original network~$G$ encoding the metric space. 
The cars and drivers will form two flows~$\fc$ and $\fd$ through~$G^T$ which are coupled in the sense that on those arcs $a\in A_L$ used for moves of convoys,
we have the condition $\fc(a)\leq \capd \cdot \fd(a)$ reflecting the dependencies between the two flows.

\paragraph*{Time-expanded network~$G^T = (V_T, A_T)$.}
The node set $V_T$ is constructed as follows: 
for each station and the depot~$v\in V$ and each time point~$t$ in the given time horizon $[0, T]$, there is a node $(v,t) \in V_T$ which represents station~$v$ at time~$t$. 
The arc set $A_T = A_H \cup A_L$ of~$G^T$ is composed of two subsets:
\begin{itemize}
\item $A_H$ contains, for each station~$v\in V$ of the original network and each $t \in \{ 0, 1, \dotsc, T-1 \} $, the \emph{holdover arc} connecting~$(v,t)$ to~$(v,t+1)$. 
\item $A_L$ contains, for each edge $(v,v')$ of~$G$ and each point in time~$t\in T$ such that $t + \dist(v,v') \leq T$, the \emph{relocation arc} from $(v,t)$ to $(v',t + \dist(v,v'))$.
\end{itemize}
Note, that the time-expanded network $G^T$ is acyclic by construction.

\paragraph*{Flows in~$G^T$.}

On the time-expanded network~$G^T$, we define two different flows, the car flow $\fc$ and the driver flow $\fd$, and specify the capacities as well as the costs for each arc with respect to both flows.

A flow on a relocation arc corresponds to a move in a tour, i.e., some cars are moved by drivers in a convoy from station~$v$ to another station~$v'$.
Hereby, the stations can be used to pick up or to drop cars, or simply to transit a node (when a driver/convoy passes the station(s) on its way to another station).
A relocation arc from $(v,t)$ to $(v', t + \dist(v,v'))$ has infinite capacity for the drivers.
However, in order to ensure that cars are moved only by drivers and only in convoys of capacity~$\capd$, we require that 
 $ \fc(a)\leq\capd \cdot \fd(a) \text{ for all $a \in A_L$.}$
Thus, the capacities for $\fc$ on the relocation arcs are not given by constants but by a function, 
see (\ref{eq: static: min-cost flows: ilp: 15}).
Each relocation arc~$a = ((v,t), (v', t + \dist(v, v')))$ corresponding to edge~$(v,v')$ has cost $\cd(a) := \dist(v,v')$.

A flow on a holdover arc corresponds to cars/drivers remaining at the station in the time interval~$[t,t+1]$.
Therefore, the capacity of all holdover arcs with respect to flow~$\fc$ is set to~$\capacity(v)$, 
see (\ref{eq: static: min-cost flows: ilp: 14}), 
whereas there is no capacity constraint for~$\fd$ on such arcs.

To correctly initialize the system, we use the nodes $(v,0) \in V_T$ as sources for both flows and set their 
balances accordingly to the initial number of cars at station~$v$ and time~$0$ and locate the drivers at the depot $v_0$, 
see (\ref{eq: static: min-cost flows: ilp: 10}).
For all internal nodes $(v, t) \in V_T$ with $0 < t < T$, we use normal flow conservation constraints, 
see (\ref{eq: static: min-cost flows: ilp: 12}) and (\ref{eq: static: min-cost flows: ilp: 13}).
To ensure that the destination state is reached and each driver returns to the depot, we use as sinks the nodes $(v, T)$, $v \in V$,
for the car flow and the node $(v_0, T)$ for the driver flow, and set their balances accordingly to $\z^T$ resp.~to the number of drivers, 
see (\ref{eq: static: min-cost flows: ilp: 11}).

The objective is to minimize the total tour length, 
see (\ref{eq: static: min-cost flows: ilp: 1}).
\begin{subequations}\label{eq: static: min-cost flows: ilp}
  \begin{flalign}
    \min& \,
\sum_{a\in A_L} \cd(a) \fd(a)\label{eq: static: min-cost flows: ilp: 1}\\
    &\sum_{a\in \delta^-(v,0)} \fc(a) = z_v^0, \sum_{a\in \delta^-(v_0, 0)} \fd(a) = k && \forall (v,0) \in V_T\label{eq: static: min-cost flows: ilp: 10}\\
    &\sum_{a\in \delta^+(v,T)} \fc(a) = z_v^T, \sum_{a\in \delta^+(v_0,T)} \fd(a) = k && \forall (v,T) \in V_T\label{eq: static: min-cost flows: ilp: 11}\\
    &\sum_{a\in \delta^-(v,t)} \fc(a) = \sum_{a\in \delta^+(v,t)} \fc(a) && \hspace{-1.6cm} \forall (v,t) \in V_T, 0 < t < T\label{eq: static: min-cost flows: ilp: 12}\\
    &\sum_{a\in \delta^-(v,t)} \fd(a) = \sum_{a\in \delta^+(v,t)} \fd(a) && \hspace{-1.6cm} \forall (v,t) \in V_T, 0 < t < T\label{eq: static: min-cost flows: ilp: 13}\\
    &0\leq\fc(a) \leq \capacity(v) && \hspace{-2.3cm} \forall a =[(v,t), (v,t+1)] \in A_H\label{eq: static: min-cost flows: ilp: 14}\\
    & \fc(a) \leq \capd \cdot \fd(a) && \forall a \in A_L\label{eq: static: min-cost flows: ilp: 15}\\
    &\fc,\fd \text{ integer},                                \label{eq: static: min-cost flows: ilp: 16}
  \end{flalign}
\end{subequations}
where $\delta^-(v,t)$ denotes the set of outgoing arcs of $(v,t)$, and $\delta^+(v,t)$ denotes the set of incoming arcs of $(v,t)$.
Note, due to the flow coupling constraints~\eqref{eq: static: min-cost flows: ilp: 15}, the above constraint matrix
is not totally unimodular (as in the case of uncoupled flows).
This reflects that the problem is~\NPhard.

Finally, the flows in the time-expanded network have to be interpreted as a transportation schedule.
Hereby, car and driver flows on relocation arcs clearly correspond to moves.
The outcome is a preemptive transportation schedule which is feasible since all dependencies over time are properly respected by the flow conservation constraints.
This implies:

\begin{theorem}
The optimal solution of system \eqref{eq: static: min-cost flows: ilp} corresponds to a preemptive transportation schedule with minimal total tour length for the Static Relocation Problem $(M,\z^0,\z^T,k,L)$.
\end{theorem}

\section{Lifted Flows in Aggregated Networks}
\label{sec: aggregated network}

The computation times for computing an exact solution by solving the integer linear program~\eqref{eq: static: min-cost flows: ilp: 1}--\eqref{eq: static: min-cost flows: ilp: 16}
are extremely high even for small instances (see Table~\ref{tab: computational results}).
This motivates the research for heuristics which compute a good feasible solution within a reasonable time.
Therefore, we describe in this section a heuristic approach
of lifted flows in aggregated networks (\LiftFlow) to solve the Static Relocation Problem $(G,\z^0,\z^T,k,L)$, 
where $G$ is the complete weighted graph $G = (V_O \cup V_U \cup \{v_0\}, E, d)$ containing the overfull stations $V_O$ (with $z_i^0 > z_i^T$),
the underfull stations $V_U$ (with $z_i^0 < z_i^T$), a depot $v_0$, all connections $E$ between them and distances $d \colon E \to \NN$.

The approach \LiftFlow\ is performed in two steps.
Firstly, we construct a weighted complete bipartite graph and find a flow that starts and ends in the depot, while passing overfull and underfull stations, minimizing the costs.
Each arc carrying a flow corresponds to a move between two stations.
Secondly, we compute possible precedences between moves and from that we construct tours (with preemption) for all convoys.

\paragraph*{First step: Flows in aggregated networks.}

In the first step, we construct an aggregated network and solve a min-cost flow problem on this network.
Arcs carrying positive flows then correspond to the movement of drivers and cars in the network.
The aggregated network is a directed, weighted graph $G^A = (V_A, A_A, w)$, based on the graph~$G$.

The node set $V_A = V_S \cup V_O \cup V_U \cup V_D$ is composed by the following nodes:
$V_S$ and~$V_D$ contain the depot,~$V_O$ contains the set of overfull stations,~$V_U$ contains the set of underfull stations.

The arc set $A_A = A_{SO} \cup A_O \cup A_U \cup A_{OU} \cup A_D$ is composed of several subsets:
\begin{itemize}
 \item the set of \emph{start arcs} $A_{SO} = \left\{ (v_0, v_o): v_0 \in V_S, v_o \in V_O \right\}$ connecting the depot to the overfull stations,
 \item the set of \emph{overfull arcs} $A_O = \left\{ (v_o,v_o'), (v_o',v_o):v_o,v_o' \in V_O \right\}$ connecting all overfull stations,
 \item the set of \emph{connection arcs} $A_{OU} = \left\{ (v_o,v_u), (v_u,v_o): v_o \in V_O, v_u \in V_U \right\}$ connecting overfull and underfull stations,
 \item the set of \emph{underfull arcs} $A_U = \left\{ (v_u,v_u'), (v_u',v_u):v_u,v_u' \in V_U \right\}$ connecting all underfull stations, and
 \item the set of \emph{sink arcs} $A_D = \left\{ (v_u, v_0): v_u \in V_U, v_0 \in V_D \right\}$.
\end{itemize}
The set containing all overfull, connection and underfull arcs is denoted by $A_R := A_O \cup A_{OU} \cup A_U$.
For an arc $a = (v, v') \in A_A$ the arc weights $w(a) := d(v, v')$ correspond to the distances.

On the aggregated network~$G^A$, we define two different flows, the car flow~$\fc$ and the driver flow~$\fd$, and specify the capacities as well as the costs for each arc with respect to both flows.
Flow on an overfull, connection or underfull arc corresponds to a move in a tour, i.e., some cars are moved by drivers in a convoy from station~$v$ to another station~$v'$.
In order to ensure that cars are moved only by drivers and that the convoy capacity~$\capd$ is not exceeded, we require that $\fc(a) \leq \capd \cdot \fd(a)$
holds for all relocation arcs $a \in A_R$, see~\eqref{eq: agg: ref: 15}.

To correctly initialize the system, we use the depot $v_0 \in V_S$ as source for the driver flow, see (\ref{eq: agg: ref: 12}), and 
the nodes $v \in V_O$ as sources for the car flow and set their balances accordingly to the numbers of cars that have to be picked up at station~$v$,
i.e., $\bc^+(v) := \max \{\z^0_v  - \z^T_v, 0 \} \geq 0$, see (\ref{eq: agg: ref: 10}).
Equalities~\eqref{eq: agg: ref: 13a} and~\eqref{eq: agg: ref: 13} are the flow conservation constraints. 
The nodes of underfull station $v \in V_U$ are the destinations of the cars.
So we set their balances accordingly to the number of cars that have to be dropped at station~$v$, i.e., $\bc^-(v) := \min \{\z^0_v  - \z^T_v, 0 \} \leq 0$, see~\eqref{eq: agg: ref: 11}.
Finally, the sink~$v_0 \in V_D$ is the destination of the $k$ drivers, see~\eqref{eq: agg: ref: 12}.
We consider a min-cost flow problem where we intend to balance all stations with minimal costs~\eqref{eq: agg: ref: 1}.
\begin{subequations} \label{eq: agg: ref}
  \begin{align}
    \min \, &   \sum_{a\in A_A} w(a) \fd(a)      \label{eq: agg: ref: 1}\\
    & \sum_{a\in \delta^-(v)} \fc(a) - \sum_{a\in \delta^+(v)} \fc(a) = \bc^+(v) &&     \text{for all $v \in V_O$}                        \label{eq: agg: ref: 10}\\
    & \sum_{a\in \delta^-(v)} \fc(a) - \sum_{a\in \delta^+(v)} \fc(a) = \bc^-(v) &&     \text{for all $v \in V_U$}                        \label{eq: agg: ref: 11}\\
    & \sum_{a\in \delta^-(v_0)} \fd(a) = \sum_{a \in \delta^+(v_0)} \fd(a) = k   &&     \text{$v_0 \in V_S, v_0 \in V_D$}                 \label{eq: agg: ref: 12}\\
    & \sum_{a\in \delta^-(v)} \fd(a) - \sum_{a\in \delta^+(v)} \fd(a) = 0        &&     \text{for all $v \in V_O\cup V_U$}                \label{eq: agg: ref: 13}\\
    & \sum_{a\in \delta^-(v)} \fc(a) - \sum_{a\in \delta^+(v)} \fc(a) = 0        &&     \text{for all $v \in V_O\cup V_U$}                \label{eq: agg: ref: 13a}\\
    & 0 \leq\fc(a) \leq \capd \cdot \fd(a)                                       &&     \text{for all $a \in A_R$}                        \label{eq: agg: ref: 15}\\
    & \fc,\fd \text{ integer},                                                                                                            \label{eq: agg: ref: 17}
  \end{align}
\end{subequations}
Note, due to constraints~\eqref{eq: agg: ref: 15}, the above constraint matrix is again not totally unimodular.

\paragraph*{Second step: Compute transportation schedule.}

Next, we describe how to compute a transportation schedule from the solution of the first step.

For that, we first compute ``pre-moves'', which have the origin and destination stations as well as the load of a move, but no times.
With these pre-moves we compute ``pre-tours'' as sequences of ``pre-moves''.
Formally, a \emph{pre-move}~$\pmove = (v, v', x)$ is a 3-tuple, where~$v = \orig(\pmove)$ is the origin station,
$v' = \dest(\pmove)$ the destination station and $x = \mloadd(\pmove)$ the load of the pre-move; a \emph{pre-tour} is a sequence of pre-moves.

Let $\fc^A$ be the car flow and $\fd^A$ the driver flow computed by the integer linear program~\eqref{eq: agg: ref} in the first step.
Then every arc $a = (v, v') \in A_A$ with $\fd^A(a) > 0$ corresponds to a pre-move $\pmove_a = (v, v', x)$, with $x \leq \fc^A(a)$.
All pre-moves have to be assigned to a pre-tour in a feasible order, i.e., the destination station of each pre-move has to be equal to the origin station of the successor pre-move.
This can be done by searching for paths within the aggregated network.
However, in general, there is not only a unique path within the aggregated network leading to several possible pre-tours.
Note that the aggregated network is not cycle-free.
Thus, there can be isolated cycles in the solution.
For this paragraph, let us assume that there are no such isolated cycles; a strategy to detect and handle isolated cycles is presented in the next paragraph.

Since we consider the preemptive situation, it is possible that there are precedences between different tours.
We define a precedence relation between pre-moves in an analog way to the definition of precedences between moves (see Section~\ref{Sec_Modeling}).
For two different pre-tours $\ptourd$ and $\ptourd'$ there exists a potential precedence between
two pre-moves $\pmove_i \in \ptourd$ and $\pmove'_j \in \ptourd'$ ($\pmove_i$ precedes $\pmove'_j$) if
\begin{itemize}
  \item the destination station $v_o$ of $\pmove_{i}$ is an overfull station, $\pmove_i$ drops cars at $v_o$, and
    $\pmove'_j$ picks up cars at $v_o$; or
  \item the origin station $v_u$ of $\pmove_{j}'$ is an underfull station, $\pmove_j'$ picks up cars at $v_u$, and
    $\pmove_i$ drops cars at $v_u$.
\end{itemize}
A pre-tour without precedences to another tour, can be directly transformed to a tour, by computing the departure and arrival times of each \mbox{(pre-)}move,
and then assigning all moves to a driver.
Otherwise, there is a preemptive situation which means that one convoy transports cars to a station and another convoy picks up these cars afterwards.
In order to ensure that the cars are dropped before they are picked up, we possibly have to add additional waiting moves to the final tour.
For that we construct a \emph{precedence relation} between pre-moves.

In some rare cases, the precedence relation is not acyclic.
In this situation, we add an additional constraint to the integer linear program~\eqref{eq: agg: ref: 1}--\eqref{eq: agg: ref: 17}, to receive a new solution.
For that we add
\[
  \sum_{a\in A_A} w(a) \fd(a) > \sum_{a\in A_A} w(a) \fd^A(a)
\]
to the constraints~\eqref{eq: agg: ref: 10}--\eqref{eq: agg: ref: 17} and recompute the steps above.
In fact, in our set of randomly generated test-instances, this case never occurred (see Table~\ref{tab: computational results}),
and we do not expect them to occur often in practice.

Thus, let us consider an acyclic precedence relation containing precedences between different pre-tours.
The arrival and departure times of the moves are derived from the distances between the origin and destination stations of the pre-moves.
When the departure time of a move is computed from a pre-move, having a precedence relation to a pre-move in another pre-tour then
we have to compute the arrival time of the preceding pre-move before, in order to be able to compute the waiting time.
If several pre-moves $\pmove_1, \dotsc, \pmove_\lambda$ precede a pre-move $\pmove$ then, in general, $\pmove$ does not need to wait for all preceding pre-moves but only
until there are enough cars at $\origin(\pmove)$.
For that, we compute a minimal set $S \subseteq \{\pmove_1, \dotsc, \pmove_\lambda\}$ by solving a minimum matching problem on
the complete bipartite graph $(\{\pmove_1, \dotsc, \pmove_\lambda\} \cup \{\pmove\}, E, w)$, where
the arc weight $w(a)$ of $a = (\pmove_j, \pmove)$ corresponds to $\mloadd(\pmove_j)$.
Hereby, the sum of the weights of the selected arcs and the number of cars at $\origin(\pmove)$ must be at least $\mloadd(\pmove)$.
The waiting time for~$\pmove$ is then induced by the latest arrival time of all $\pmove_j \in S$.

After all waiting times have been computed, we construct a transportation schedule from the (waiting) moves.

\paragraph*{Handling cycles in the lifted flows.}

Finally, we describe how isolated cycles in the lifted flows can be handled.
Let $v_1 \dotsm v_\lambda v_1$ be an isolated cycle with $\fd^A(v_1, v_2) = \dotsm = \fd^A(v_{\lambda - 1}, v_\lambda) = \fd^A(v_\lambda, v_1)$.
Furthermore, let the cycle be ordered so that $v_1$ is an overfull station and $v_\lambda$ is an underfull station and so that $\fc^A(v_\lambda, v_1) = 0$.
Note, the conservation constraints ensure that isolated cycles contain overfull and underfull stations.
Furthermore, the existence of an arc from an underfull to an overfull station with $\fc^A(v_\lambda, v_1) = 0$ is ensured by the minimality of the car flows.
Since the cycle is isolated, it holds by definition $\fd^A(v, v_j) = \fd^A(v_j, v) = 0$ for all $v \in V_A \setminus \{ v_1, \dotsc, v_\lambda \}$.

Since $v_1$ is an overfull station and $v_\lambda$ an underfull station, we can consider the path $v_1 \dotsm v_\lambda$ as a subsequence of pre-moves.
Precedences within this subsequence are then handled as described in the previous paragraph.

Let $\tourd$ be a tour from the transportation schedule computed in Step~2, and let~$\move$ be the last move from~$\tourd$.
Since there are no cars transfered
from~$v_\lambda$ to~$v_1$, we can modify~$\tourd$ by removing~$\move$ and adding a move from $\orig(\move)$ to $v_1$,
as well as the moves induced from the path $v_1 \dotsm v_\lambda$.
Finally, a move from~$v_\lambda$ to the depot~$\dest(\move)$ is added.

Since we can handle precedences between pre-moves and the case of isolated cycles in the two flows, we obtain:
\begin{theorem}
 The approach \LiftFlow\ computes a feasible (possibly preemptive) transportation schedule for the Static Relocation Problem.
\end{theorem}

\paragraph*{Optimal solutions and lower bounds.}

Under certain conditions, the algorithm~\LiftFlow\ computes an optimal solution for the Static Relocation Problem or provides a lower bound for an optimal solution.

\begin{theorem} \label{thm: static: aggregated network: an 2 ten}
Let $(G,\z^0,\z^T,k,L)$ be a Static Relocation Problem, $\fc^A$ and $\fd^A$ optimal flows in the aggregated network $G^A$, and $\sched$ be the resulting transportation schedule. 
If $\fc^A$ and $\fd^A$ satisfy that
  \begin{enumerate}
    \item\label{thm: static: aggregated network: an 2 ten: 1} there are no cycles in the precedence graph,
    \item\label{thm: static: aggregated network: an 2 ten: 2} there are no isolated cycles in the flows,
    \item\label{thm: static: aggregated network: an 2 ten: 4} the capacities of the stations are sufficiently large,
    \item\label{thm: static: aggregated network: an 2 ten: 5} the time horizon is sufficiently large, and
    \item\label{thm: static: aggregated network: an 2 ten: 3} there exists an optimal solution so that no balanced station is used as preemption station,
  \end{enumerate}
then $\sched$ is an optimal solution for~$(G,\z^0,\z^T,k,L)$. 
If only condition 5 is fulfilled, then $\sum_{a \in A_A} w(a) \fd^A(a)$ is a lower bound for the tour length of an optimal solution for~$(G,\z^0,\z^T,k,L)$.
\end{theorem}

\section{Computational Results}

Both, the heuristic approach \LiftFlow\ and the exact approach, have been tested on randomly generated instances (with 20--80 over-/underfull stations, 50--150 stations in total, convoy capacities 5 and 10, and 10--30 drivers). 
The stations are randomly distributed on a plane and the distances between two closest stations (w.r.t.~the Euclidean metric) are kept as rounded integers in the graph.
Hereby, we ensure that the graph is connected.
The time horizons are set to 100 and 150.
Note that the size of these instances corresponds to small car- or bikesharing systems or to clusters of larger systems, as in \cite{SHH-2013}. 
The combinatorial part of the algorithm \LiftFlow\ has been implemented in Java 1.6, Gurobi 5.6 is used for solving the integer linear program (short:~ILP) in the first step of the algorithm.
Gurobi 5.6 is also used to solve the ILPs for the exact approach.
The tests have been run on a server with a total of 160 Intel Xeon E7-8870 cores, each clocked at 2.40GHz, and with 1 TB RAM.
The number of threads for solving an ILP is restricted to 16, the implementation of \LiftFlow\ is single-threaded.

In the heuristic approach \LiftFlow, most of the computation time is used for solving the min-cost flows in the aggregated network.
Hereby, we observed that after 20 minutes computation time for the first step, the objective function value rarely improved,
see Table~\ref{tab: computational results} showing the results of solving the ILP in the first step of \LiftFlow\ after 20 minutes and 4 hours.
That the results are better in some cases after 20 minutes than after 4 hours is due to a different number of isolated cycles.
The computation time for computing the transportation schedule from the flows in the second step is 40 seconds for all tours. 
Due to the small impact, these run times are not listed in the table.

For the exact approach, the optimal solution could not be found for any test instance within the given time limit.
Hereby, we state the best found solution values from a set of different optimization parameters on solving the ILPs.
The time limits have been set to at least 4 hours (once we have let the computation running for nearly 16 days).
In the first part of the algorithm \LiftFlow, the optimal solution was found in 6 instances within the time limits.
In eight (resp.~seven) examples, \LiftFlow\ computes tours exceeding the given time horizon leading to a non-feasible transportation schedule.
However, most of the times, these transportation schedules can be modified by splitting the concerning tours into two (or more) tours so that they fit in the time horizon.
This proposal works if there are enough drivers available.

\begin{table}[ht]
\centering
 \caption{
    This table shows the computational results for several test instances of the algorithm \LiftFlow\ (the time limit was set to 20 minutes and 4 hours)
    in comparison to the value found by solving the ILP (the time limit was set at least 4h).
    For every parameter set we created two different test instances ($a$ and $b$).
    In this table, the following parameters and results are shown:
    the name of the test instance $a$ or $b$ (1st column),
    the total amount of stations (2nd column) and the number of overfull and underfull stations (3rd column).
    Furthermore, it shows the considered time horizon $T$, the server capacity~$\capd$,  the number of drivers $k$,
    the total tour length LF found by \LiftFlow\ (after 20 minutes and 4 hours, respectively),
    a lower bound LB (see Theorem~\ref{thm: static: aggregated network: an 2 ten} assuming condition~\ref{thm: static: aggregated network: an 2 ten: 3} holds),
    the duality gap between \LiftFlow\ (20min) and LB in percent, by the ILP solver after 4 hours and the duality gap between ILP and LB, 
    and by an algorithm with an approximation factor \REOPT\ (see~\cite{LAGOS2013} for details) and the median runtime t in seconds of several runs.
    Values marked with an asterisk (*) contain a tour exceeding the time horizon, i.e., it is not a feasible transportation schedule;
    the ILP within \LiftFlow\ has been solved optimally when values are marked with a plus (+), otherwise the lower bound is computed by the ILP solver;
    the cells marked with a hyphen (-) indicate that no solution was found within the time limit.}
 \label{tab: computational results}
{\scriptsize 
\begin{tabular}{c|c|c|c|c|c|c|c|l|c|c|c|c|c}
instance&stations&$\pm$stations& T   &$\capd$ & k   & LF (20min)       & LF (4h)       & LB      & GAP\%  & ILP & GAP\% & \REOPT & t(s) \\
\hline
$a$     & 050    & 10/10       & 100 & 05     & 010 & 272*             & 272*          & $232^+$ & 14.7   & 312 & 25.6  &  372   & 2 \\
$b$     & 050    & 10/10       & 100 & 05     & 010 & 315*             & 315\;\,       & $238^+$ & 24.4   & 374 & 36.4  &  -     & - \\
$a$     & 050    & 10/10       & 100 & 10     & 010 & 255*             & 255*          & $221^+$ & 13.3   & 287 & 23.0  &  328   & 3 \\
$b$     & 050    & 10/10       & 100 & 10     & 010 & 254\;\,          & 254*          & $188^+$ & 23.3   & 292 & 35.6  &  -     & - \\
$a$     & 050    & 20/20       & 100 & 05     & 020 & 319\;\,          & 310*          & 285     &  8.7   & 404 & 28.0  &  438   & 85 \\
$b$     & 050    & 20/20       & 100 & 05     & 020 & 378*             & 373\;\,       & $334^+$ & 11.6   & 467 & 28.5  &  519   & 78 \\
$a$     & 050    & 20/20       & 100 & 10     & 020 & 291\;\,          & 291\;\,       & 245     &  6.5   & 369 & 26.3  &  410   & 70 \\
$b$     & 050    & 20/20       & 100 & 10     & 020 & 282\;\,          & 292\;\,       & $262^+$ &  7.1   & 385 & 31.9  &  420   & 60 \\ \hline
$a$     & 100    & 30/30       & 100 & 05     & 030 & 533\;\,          & 513\;\,       & 435     & 15.4   & 736 & 38.7  &  -     & - \\
$b$     & 100    & 30/30       & 100 & 05     & 030 & 448\;\,          & 432\;\,       & 361     & 16.5   & 616 & 39.3  &  673   & 482 \\
$a$     & 100    & 30/30       & 100 & 10     & 030 & 396*             & 412*          & 322     & 11.4   & 510 & 31.2  &  -     & - \\
$b$     & 100    & 30/30       & 100 & 10     & 030 & 327*             & 327*          & 269     & 12.5   & 470 & 39.1  &  610   & 496 \\
$a$     & 100    & 40/40       & 100 & 05     & 030 & 536*             & 539*          & 471     &  8.2   & 765 & 35.7  &  -     & - \\
$b$     & 100    & 40/40       & 100 & 05     & 030 & 469\;\,          & 469\;\,       & 367     & 16.0   & 599 & 34.2  &  668   & 1391 \\
$a$     & 100    & 40/40       & 100 & 10     & 030 & 434*             & 434\;\,       & 340     &  9.7   & 585 & 33.0  &  -     & - \\
$b$     & 100    & 40/40       & 100 & 10     & 030 & 388\;\,          & 417\;\,       & 285     & 14.4   & 473 & 29.8  &  614   & 713 \\ \hline
$a$     & 150    & 50/50       & 150 & 05     & 030 & 686\;\,          & 690\;\,       & 535     & 17.2   & 944 & 39.5  &  -     & - \\
$b$     & 150    & 50/50       & 150 & 05     & 030 & 717\;\,          & 707\;\,       & 594     & 13.3   & 1008& 39.2  &  -     & - \\
$a$     & 150    & 50/50       & 150 & 10     & 030 & 474\;\,          & 476\;\,       & 378     &  8.6   & 675 & 35.6  &  -     & - \\
$b$     & 150    & 50/50       & 150 & 10     & 030 & 533\;\,          & 534\;\,       & 415     & 13.5   & 812 & 43.1  &  -     & - \\ \hline \hline
\multicolumn{8}{ l }{average}                                                          &         & 13.3   &     & 33.7  
\end{tabular}}
\end{table}

\vspace{-0.2cm}

\section{Conclusion}

In this paper, we considered the Static Relocation Problem $(G,\z^0,\z^T,k,L)$, where
tours for $k$ drivers have to be computed in a graph $G$, whereas the maximal length of the tours must fit into a given time horizon~$T$.
Hereby, we compute preemptive transportation schedules: a car can be transported in one convoy from its origin to an intermediate station, and from there by another convoy to its destination.
In order to have an exact solution we construct a time-expanded network $G^T$ from the original network $G$ and compute two coupled flows (a car and a driver flow)
on this network with an integer linear program.
Due to the coupling constraints, the obtained constraint matrix is not totally unimodular (as in the case of uncoupled flows).

Due to the high computational times of the exact approach and the fact that algorithm \REOPT, which computes non-preemptive transportation schedules, not always finds a feasible solution,
we developed a heuristic approach to solve the Static Relocation Problem: the algorithm \LiftFlow.
The construction of the tours by \LiftFlow\ is as follows.
In the first step, two coupled flows on a graph are computed, a driver and a car flow.
These flows serve as input for the second step, where firstly a set of pre-tours is constructed.
Afterwards, precedence relations between pre-tours are computed, and
from these precedences and the set of pre-tours, we finally compute a transportation schedule.

The heuristic approach \LiftFlow\ solves the test instances faster and, computes already after 20 minutes transportation schedules with shorter total tour lengths
than the exact approach
using time-expanded networks 
with a time limit of 4 hours (see Table~\ref{tab: computational results}).

Under certain conditions, \LiftFlow\ computes an optimal solution.
Otherwise, we receive at least a lower bound (Theorem~\ref{thm: static: aggregated network: an 2 ten}).

There are several practical and theoretical open questions according to the Static Relocation Problem.
Currently, we minimize the total tour length.
Applying the ideas of \LiftFlow\ so that the makespan is minimized is one goal for the future.
Another future goal is to improve the runtime of solving the min-cost flow in the aggregated network of the algorithm \LiftFlow.
This may be achieved by improving the lower bound from the dual solution.
Usually, every driver used gives additional costs.
Thus, it is desirable to know the minimal and/or maximal number of drivers needed in order to solve the Static Relocation Problem within the given time horizon.
To the best of our knowledge this is still an open question.
Due to the time horizon, it is possible that there does not exist a feasible solution for a given instance at all.
Having feasibility conditions is useful in two directions: to save unnecessary computation time for an algorithm and to generate test instances which can give feasible solutions.
Thus, finding feasibility conditions as well as lower bounds for the time horizon is another future goal.

\bibliographystyle{plain}
\providecommand{\availatURL}[1]{\ignorespaces \footnote{Avail. at URL
  \texttt{#1}}} \providecommand{\NP}{\textsf{NP}}
\providecommand{\bysame}{\leavevmode\hbox to3em{\hrulefill}\thinspace}
\providecommand{\MR}{\relax\ifhmode\unskip\space\fi MR }
\providecommand{\MRhref}[2]{%
  \href{http://www.ams.org/mathscinet-getitem?mr=#1}{#2}
}
\providecommand{\href}[2]{#2}


\begin{thebibliography}{1}

\bibitem{Ball+etal:handbook:95a}
M.~O. Ball, T.~L. Magnanti, C.~L. Monma, and G.~L. Nemhauser, editors.
\newblock {\em Network Models}, volume~7 of {\em Handbooks in Operations
  Research and Management Science}.
\newblock Elsevier Science B.V., Amsterdam, 1995.

\bibitem{Benchimol+etal:RAIRO}
M.~Benchimol, P.~Benchimol, B.~Chappert, A.~de~la Taille, F.~Laroche,
  F.~Meunier, and L.~Robinet.
\newblock Balancing the stations of a self service “bike hire” system.
\newblock {\em RAIRO - Operations Research}, 45:37--61, 0 2011.

\bibitem{do-cmc2013}
D.~Chemla, F.~Meunier, and R.~Wolfler Calvo.
\newblock Bike sharing systems: {S}olving the static rebalancing problem.
\newblock pages 120--146, 2013.

\bibitem{cirrelt-CMR-2012}
C.~Contardo, C.~Morency, and L-M. Rousseau.
\newblock Balancing a dynamic public bike-sharing system.
\newblock Technical Report~9, CIRRELT, 2012.
\newblock \url{https://www.cirrelt.ca/DocumentsTravail/CIRRELT-2012-09.pdf}.

\bibitem{EDGC:2012:PCS}
M.~EL-Zaher, B.~Dafflon, F.~Gechter, and J.-M. Contet.
\newblock Vehicle platoon control with multi-configuration ability.
\newblock {\em Procedia Computer Science}, 9(0):1503--1512, 2012.

\bibitem{LAGOS2013}
S.~O. Krumke, A.~Quilliot, A.~K. Wagler, and J.-T. Wegener.
\newblock Models and algorithms for carsharing systems and related problems.
\newblock {\em Electronic Notes in Discrete Mathematics}, 44(0):201 -- 206,
  2013.

\bibitem{KQWW:2014:LNCS}
S.~O. Krumke, A.~Quilliot, A.~K. Wagler, and J.-T. Wegener.
\newblock Relocation in carsharing systems using flows in time-expanded
  networks.
\newblock In {\em Experimental Algorithms}, volume 8504 of {\em LNCS}, pages
  87--98. Springer, 2014.

\bibitem{EvoCOP:HPHR-2013}
M.~Rainer-Harbach, P.~Papazek, B.~Hu, and G.~R. Raidl.
\newblock Balancing bicycle sharing systems: A variable neighborhood search
  approach.
\newblock In {\em EvoCOP}, pages 121--132, 2013.

\bibitem{SHH-2013}
J.~Schuijbroek, R.~Hampshire, and W.-J. van Hoeve.
\newblock Inventory rebalancing and vehicle routing in bike sharing systems.
\newblock 2013.
\newblock working paper.

\end{thebibliography}

\end{document}